\documentclass[twocolumn,journal]{IEEEtran}
\usepackage[T1]{fontenc}
\usepackage[latin9]{inputenc}
\usepackage{fancybox}
\usepackage{calc}
\usepackage{graphicx}
\usepackage[unicode=true,
 bookmarks=true,bookmarksnumbered=true,bookmarksopen=true,bookmarksopenlevel=1,
 breaklinks=false,pdfborder={0 0 0},pdfborderstyle={},backref=false,colorlinks=false]
 {hyperref}
\hypersetup{pdftitle={Your Title},
 pdfauthor={Your Name},
 pdfpagelayout=OneColumn, pdfnewwindow=true, pdfstartview=XYZ, plainpages=false}

\makeatletter
\usepackage[caption=false,font=footnotesize]{subfig}
\usepackage{diagbox}
\usepackage{algorithm}
\usepackage{algorithmic}
\usepackage{color}
\usepackage{amsmath}
\usepackage{bm}

\makeatother

\begin{document}
\title{Deep Learning for Near-Field XL-MIMO Transceiver Design: Principles
and Techniques}

\author{Wentao~Yu,~\IEEEmembership{Student Member,~IEEE,}~Yifan~Ma,~\IEEEmembership{Student Member,~IEEE,}~Hengtao~He,~\IEEEmembership{Member,~IEEE,}
Shenghui~Song,~\IEEEmembership{Senior Member,~IEEE,} Jun~Zhang,~\IEEEmembership{Fellow,~IEEE,}
and~Khaled~B. Letaief,~\IEEEmembership{Fellow,~IEEE}\thanks{The authors are with the Dept. of ECE, The Hong Kong University of Science and Technology, Hong Kong. \textit{(Corresponding author: Hengtao He.)} }}

%\markboth{Preprint. Submitted to IEEE. }{}
\markboth{Accepted by IEEE Communications Magazine}{}
\maketitle
\begin{abstract}
Massive multiple-input multiple-output (MIMO) has been a critical enabling technology in 5th generation (5G) wireless networks. With the advent of 6G, a natural evolution is to employ even more antennas, potentially an order of magnitude more, to meet the ever-increasing demand for spectral efficiency. This is beyond a mere quantitative scale-up. The enlarged array aperture brings a paradigm shift towards near-field communications, departing from traditional far-field approaches. However, designing advanced transceiver algorithms for near-field systems
is extremely challenging because of the enormous system scale, the complicated
channel characteristics, and the uncertainties in the propagation environments.
Hence, it is important to develop \textit{scalable}, \textit{low-complexity},
and \textit{robust} algorithms that can efficiently characterize and
leverage the properties of the \textit{near-field channel}. In this article, we discuss the principles and advocate two general frameworks to design deep learning-based near-field transceivers covering both iterative and non-iterative algorithms. Case studies on channel estimation and beam focusing are presented to provide a hands-on tutorial. Finally, we discuss open issues and shed light on future directions. 
\end{abstract}

\IEEEpeerreviewmaketitle{}

\section{Introduction}

As the deployment of 5G wireless networks continues
to accelerate throughout the world, we are witnessing exciting global research and development
activities to materialize 6G as the next-generation mobile
communication network. Future 6G networks will revolutionize wireless
communications from \textquotedblleft connected things\textquotedblright{}
to \textquotedblleft connected intelligence\textquotedblright , enabling
ubiquitous inter-connections between humans, things, and intelligence
within a deeply intertwined and hyper-connected cyber-physical world \cite{2022Letaief}. To realize these ambitions, stringent
requirements for the physical layer must be met, such as exceptionally
high data rates, extremely low end-to-end latency, exceedingly high
spectral efficiency, and massive connections. 

Extremely large-scale multiple-input multiple-output (XL-MIMO) systems \cite{2023Wang}, with potentially
thousands of transmit and receive antennas, were proposed as a disruptive physical layer technology to unleash the full potential
of mmWave and THz communication systems \cite{2020Faisal}. In addition
to overcoming the high propagation loss, XL-MIMO also brings other benefits, including ultra-high data rates and spectral efficiency, through enhanced spatial
multiplexing. Most importantly, the enlarged array aperture, together with the small carrier wavelength, causes an explosion in the Rayleigh
distance and
triggers a paradigm shift from traditional far-field communications
to the near-field regime. Fig. \ref{fig:A-typical-UM-MIMO} shows that the near-field region typically covers a big portion of the cell in mmWave
and THz XL-MIMO \cite{2023Yu-JSTSP}. 

Near-field propagation leads to many valuable opportunities as well as
stringent challenges for the transceiver design. The spherical wavefront provides a new degree of freedom, i.e., the distance dimension, for transmission and sensing. It allows the near-field beam to focus on an exact point rather than a rough angular
direction, which facilitates interference mitigation and thus improves
the sum-rate in multi-user communications \cite{2022Zhang}. On the other hand, the spherical wavefront also complicates the acquisition,
representation, and exploitation of the channel. 
A significantly larger codebook is
required to sparsify the near-field channel, as the codewords should sample both the angle and distance domains \cite{2023Cui}. Also, near-field XL-MIMO systems are easily subject to blockages and hybrid-field transitions
\cite{2023Yu-JSTSP}, which may lead to dynamic propagation environments. 

The challenges for near-field XL-MIMO transceiver design are three-fold,
i.e., the complicated channel characteristics, the enormous system
scale, and the uncertainties in propagation environments. Given the
recent success of deep learning (DL), in solving challenging
problems, one naturally
considers: \smallskip \smallskip

\noindent\doublebox{\begin{minipage}[t]{1\columnwidth - 2\fboxsep - 7.5\fboxrule - 1pt}%
\begin{center}
\textit{Is it possible to design DL-based transceivers that are }\textbf{\textit{scalable}}\textit{,
}\textbf{\textit{efficient}}\textit{, }\textbf{\textit{robust}}\textit{,
and can effectively exploit the near-field channel characteristics?}
\par\end{center}%
\end{minipage}}

\smallskip In this paper, we provide affirmative answers to this
question with a holistic view of principles and techniques of DL-based design for near-field XL-MIMO transceivers. In Section II, we identify the common research challenges and discuss the roles and requirements of DL in tackling them. In Section III, we establish two scalable, low-complexity, and robust model-driven DL frameworks
for near-field XL-MIMO transceivers, i.e., fixed point networks
(FPNs) and neural calibration (NC), which are respectively designed
for iterative and non-iterative algorithms, and analyze their advantages compared with the existing techniques. In Section IV, we present two tutorial-style case studies on channel estimation and beam focusing. Lastly, we discuss future directions and conclude the paper in Sections V and VI, respectively. 

\textit{Related works:} Different from previous works that focus on the application of DL to general communications and signal processing problems, this work particularly targets the challenges in near-field XL-MIMO and tackles the limitations of prior methods \cite{2020Bjornson,2021Monga,2019He}. Different from \cite{2020Bjornson}, we will critically examine the limited scalability of unstructured neural networks in near-field XL-MIMO and introduce structured frameworks as a better alternative. Unlike \cite{2021Monga} and \cite{2019He}, we will point out the shortcomings of deep unfolding and improve its efficiency and scalability in large-scale near-field systems by enhancing the training and inference processes, leading to better convergence and generalization. Both iterative and non-iterative algorithms will be discussed. We also complement these discussions with extensive simulation results, and provide tutorials and research directions to inspire future endeavors. 

\section{Research Challenges and the Roles of DL}

We first introduce the common research challenges and then discuss the roles and requirements of DL in the design of near-field XL-MIMO transceivers. 

\subsection{Research Challenges}

\subsubsection{Complicated Channel Characteristics}

\begin{figure}[t]
\centering{}\includegraphics[width=8.5cm]{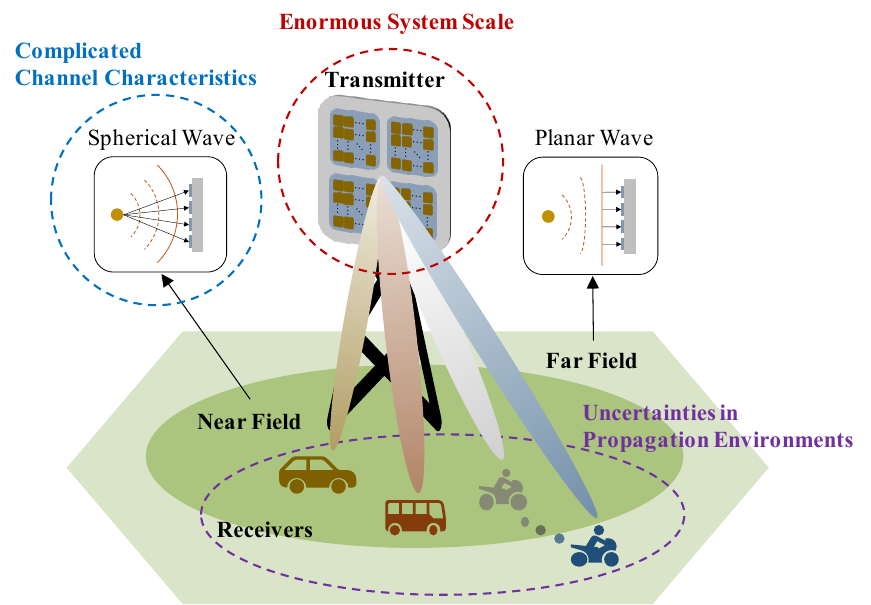}\caption{A typical XL-MIMO system involving near-field propagation. 
\label{fig:A-typical-UM-MIMO}}
\end{figure}

The enlarged dictionary matrix required to represent the near-field channel is a crucial challenge of near-field XL-MIMO transceiver design. As shown in Fig. \ref{fig:A-typical-UM-MIMO},
in far-field systems, the angles of arrival (AoAs) at different antennas
and the phase differences between adjacent antenna elements
are both identical. However,
the AoAs in near-field systems vary at different antennas,
jointly determined by the reference AoA and the propagation distance. A significantly larger dictionary that samples both the angular and distance domains is
required to sparsify the near-field channel, which causes an excessive
overhead. For example, the dimension
of the near-field dictionary utilized in beam training and channel estimation
is several times that of the far-field counterpart \cite{2023Liu}.
This will lead to a prohibitive surge in complexity
if the same algorithms are adopted \cite{2023Yu-JSTSP}. 
 
\subsubsection{Enormous System Scale}

The second research challenge is attributed to the enormous system scale.
In near-field XL-MIMO systems, signal processing and optimization algorithms
should handle problems with an extremely high dimension owing to the huge number of antennas, the ever-increasing number
of supportable users, and the enlarged bandwidth. Classical algorithms demonstrate a prohibitive complexity in such systems. For example, state-of-the-art
(SoA) orthogonal approximate message passing (OAMP)-based detectors require a high-dimensional matrix inversion per iteration \cite{2022Kosasih}, This leads to a prohibitive complexity that is cubic with respect to the antenna number, which poses severe challenges
to the real-time operation of the transceivers. Thus, designing algorithms with low complexity and strong scalability is crucial. 

\subsubsection{Uncertainties in Propagation Environments}

Uncertainties in the propagation environment constitute the third challenge. Near-field XL-MIMO systems operating at mmWave and
THz bands are vulnerable to various types of blockage due to 
poor penetration. This will result in channel
and signal-to-noise ratio (SNR) fluctuations. In addition, each
user terminal may view different parts of the XL-MIMO array,
referred to as the visible region. The resultant spatial non-stationary
effect should be taken into account. Furthermore, depending
on the specific array aperture and carrier frequency, each cell may
consist of a mixture of the far- and near-field regions, termed
the hybrid-field scenario \cite{2023Yu-JSTSP}. User mobility
can result in transitions between the different fields. 

\subsection{The Roles and Requirements of DL}

In recent years, DL-based methods has gradually become an indispensable
set of tools for advanced transceiver design \cite{2020Bjornson}. Specifically, they are especially suitable to tackle the `hard to compute'
and `hard to model' problems that are prevalent in near-field XL-MIMO
systems. 

First, the complexity of the problems in XL-MIMO systems is rapidly approaching
the point that exceeds the capabilities of traditional approaches, which involve many computationally intensive operations,
e.g., matrix inversion and singular value decomposition (SVD). The well-trained neural network-based solver can offload the
complexity from the online inference to the offline training stage. In addition, as the near-field channels become increasingly complicated to model, the traditional
sparsity-based and Bayesian approaches may not efficiently capture
channel characteristics \cite{2023Yu-JSTSP}. Thanks to the universal approximation property, DL can be utilized to model the high-dimensional
channel distribution in a data-driven manner to improve the performance
of various transceiver modules. 

However, the near-field XL-MIMO channels will magnify the following
issues in designing DL-based transceivers: 

\subsubsection{Scalability and Complexity}

Since the system scale of XL-MIMO is far larger than that of legacy
systems, multiple aspects of scalability must be considered, including
sample, time, and space complexity. Sample complexity refers to the number of training samples
required to achieve satisfactory performance. Time and space complexity are both measures of the computational burden of the transceiver algorithms. A flexible tradeoff between complexity and performance should be achieved to satisfy the real-time requirements.

\subsubsection{Generalization and Adaptation}

Before the deployment of DL-based algorithms in near-field XL-MIMO transceivers,
an open challenge is how to enable the trained neural networks to adapt to different
wireless scenarios since the practical propagation environments can be highly
dynamic. Such an ability is called \textit{generalization},
which characterizes the adaptability of the DL models to previously
unseen data. Good generalization calls for accurate predictions with limited or no fine-tuning. 

\subsubsection{Theoretical Guarantees}

The lack of theoretical guarantees is another concern that
hinders the widespread deployment of DL-based transceivers.
Traditional physical layer algorithms have deeply rooted mathematical
foundations and guarantees. Nevertheless, DL-based
designs, even the model-driven ones that stem from classical algorithms,
generally lack theoretical supports. To enhance the reliability of DL-based transceivers, it is crucial to study and provide concrete theoretical supports for the general algorithmic frameworks.

\section{Algorithmic Frameworks and Enabling Techniques}

In this section, we propose two DL-based frameworks for the non-iterative and
iterative algorithms in designing near-field XL-MIMO transceivers, respectively,
which are scalable, low-complexity, robust, and have some initial theoretical supports.
We discuss their advantages through comparison with existing DL-based
algorithms \cite{2019He}, as illustrated in Table \ref{tab:Comparison-Existing-Proposed}.
\begin{table*}[t]
\caption{A Comparison between the Proposed and the Existing Algorithmic Frameworks
\label{tab:Comparison-Existing-Proposed}}
\centering{}\includegraphics[width=17.5cm]{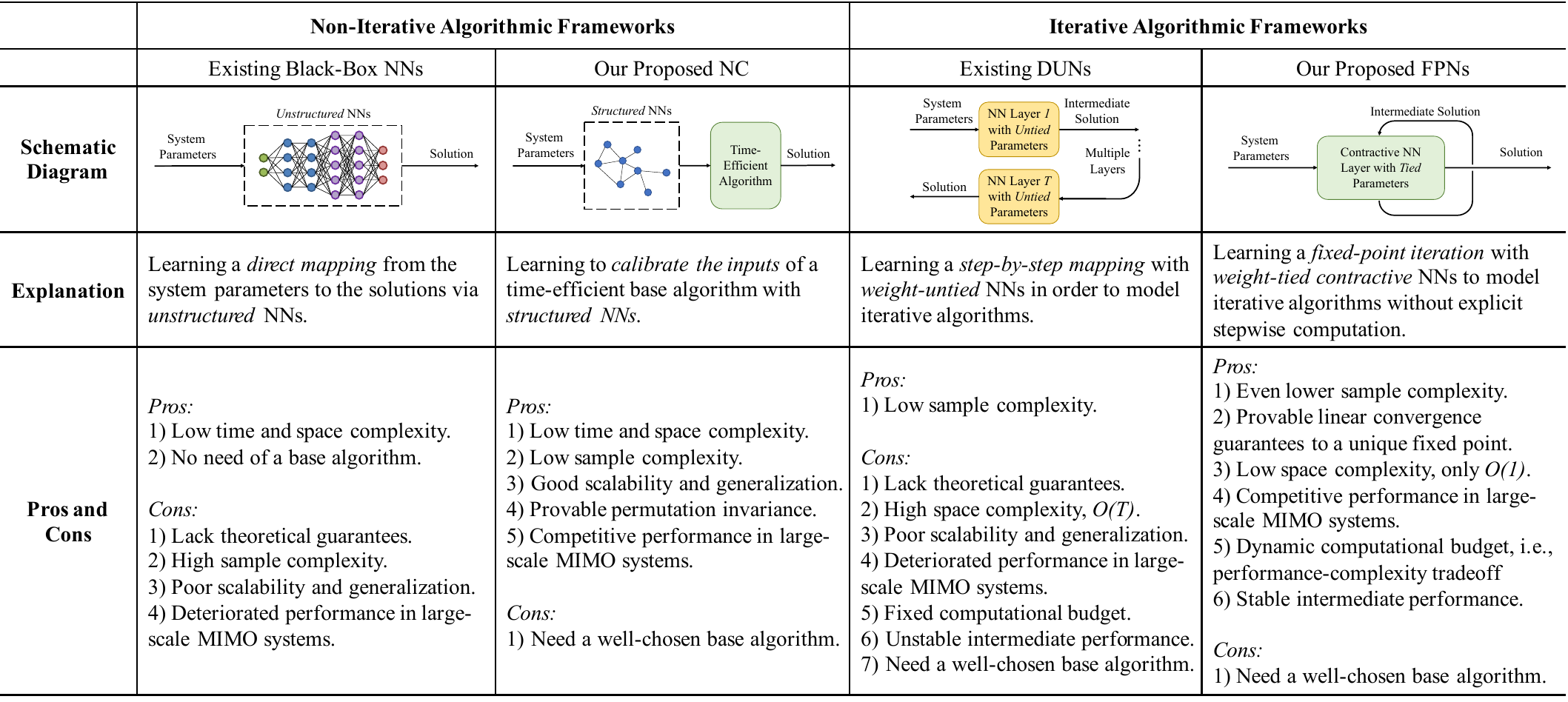}
\end{table*}

\subsection{Algorithmic Frameworks for Non-Iterative Algorithms}

\subsubsection{Existing Black-Box Neural Networks (NNs)}

Most existing works belong to the fully data-driven framework,
where black-box NNs replace classical model-based algorithms
to approximate the mapping from the system parameters
to the solutions, as illustrated in the first column
of Table \ref{tab:Comparison-Existing-Proposed}. 

Though black-box NNs can be executed in real-time, they generally
lack acceptable scalability and generalization ability when the network
size is huge. For typical black-box NNs, e.g., MLPs and convolutional neural networks
(CNNs), the underlying high-dimensional mapping is difficult to learn
in large-scale wireless networks. 
Moreover, as MLPs and CNNs have fixed input and output dimensions, they can hardly generalize if the dimensions of optimization variables differ in the training and inference stages. The NNs must be re-trained if the number of users changes, which often
happens in near-field XL-MIMO systems due to constant blockage.

\subsubsection{Proposed Neural Calibration (NC)}

To tackle the drawbacks of black-box NNs, we propose an NC framework
that is significantly more scalable and generalizable \cite{2022Ma}.
As shown in the second column of Table \ref{tab:Comparison-Existing-Proposed},
two components are involved in the proposed NC framework, i.e., a
structured NN model followed by a time-efficient algorithm as the
calibration basis. The backbone of the low-complexity algorithm is retained
while NNs are adopted to calibrate the input to improve the system
performance. With different algorithms as the basis of calibration, the NC framework is widely compatible with a variety of transceiver modules. For example, for the near-field beam focusing unit, the basis algorithm can be
selected as the low-complexity zero-forcing (ZF) or maximum ratio transmission (MRT) methods, which can both be easily differentiable and will not affect the gradient back-propagation in the training stage.

A unique structured architecture is further developed for the NN module
in the proposed NC framework. Instead of adopting an unstructured black-box
NN, the proposed method is based on the permutation
equivariance (PE) property and naturally inherits the structure of
wireless network topology. Specifically, PE refers to the characteristic
of an algorithm where its output remains invariant under any permutation
of the input elements, i.e., rearranging the order of the input elements does not change the corresponding output or result. The PE property is universal for wireless communication problems. The structured NN in the proposed NC framework satisfies the PE property, where the training process ignores the unnecessary indexing issue and focuses purely on learning the solution value. This is particularly important for the near-field XL-MIMO transceivers, whose design typically incurs non-trivial high-dimensional problems.
We will then show how the proposed NC framework addresses the existing challenges for near-field XL-MIMO transceivers:
\begin{itemize}
\item{\textbf{Scalability and Complexity:}}
With the order irrelevance PE property, the structured NN used in NC leads to parallel processing for different users/antennas. This, together with the time-efficient basis algorithm, results in a low computational complexity of NC. Furthermore, the input and output dimensions of the proposed NN are independent of the number of users/antennas, which leads to a reduced number of trainable parameters and a highly scalable transceiver design for large-scale systems.
\item{\textbf{Theoretical Guarantee:}} In the NC framework, two PE properties are proved in \cite{2022Ma}, i.e., downlink user-wise PE and uplink antenna-wise PE, which are suitable for downlink beamforming design and uplink channel estimation, respectively. Moreover, the existence of the calibration mapping is also proved in \cite{2022Ma}.
\item{\textbf{Generalization:}}
Thanks to the PE properties, the proposed NC framework can flexibly
adapt to different system sizes, e.g., the number of users or antennas,
without retraining the NNs while enjoying good performance. 
\end{itemize}

A tutorial-style case study will be further supplemented in Section \ref{subsec:Beam-Focusing}
to illustrate specific designs of the NC framework for beam focusing
in near-field XL-MIMO systems.

\subsection{Algorithmic Frameworks for Iterative Algorithms}

\subsubsection{Existing Deep Unfolding Networks (DUNs)}

DUNs were proposed as a paradigm that combines wireless domain knowledge
with DL \cite{2019He}.
In particular, a DUN aims at imitating the structure of an iterative
algorithm by converting each iteration to one layer of NN with untied
trainable weights, as shown in the third column of Table \ref{tab:Comparison-Existing-Proposed}.
DUN is a step-by-step process where the number of
layers at the training and testing stages should be identical \cite{2019He,2021Elbir}.

However, existing DUNs suffer from multiple
crucial drawbacks. The first problem lies in the poor scalability \cite{2022Ma}. The training
process of DUNs requires the storage of intermediate states
and gradients for each layer, which results in significant memory and
computational burdens. This limitation hampers scalability to large-scale problems, particularly those regarding XL-MIMO arrays. Furthermore, the complexity
of DUNs cannot be adapted at the inference stage \cite{2023Yu-JSTSP}. Whether converged or not, DUNs produce a solution after running a fixed
number of layers. The third concern pertains to the reliability issue.
Classical algorithms usually iteratively enhance the quality of the solution until reaching a fixed point. However, DUNs
abruptly halt after a pre-determined number of layers, which disagrees
with the nature of classical algorithms and lacks theoretical
guarantees. The intermediate states tend to oscillate rather than converge. Lastly, the generalization
issue also remains unsolved. DUNs suffer potential performance degradation in out-of-distribution
scenarios \cite{2023Yu-JSTSP}, which frequently occurs
in near-field XL-MIMO transceivers if the distributions of the
channel, pilot patterns, and noise in the testing environment
differ from those in the training stage.

\subsubsection{Proposed Fixed Point Networks (FPNs)}

To address the limitations of DUNs, FPNs were introduced in \cite{2023Yu-JSTSP}
as a versatile framework that constructs iterative transceiver algorithms
by utilizing the fixed point iteration of a learnable contraction mapping. This idea is originated from the fact that the outputs of
classical iterative algorithms, e.g., approximate message passing (AMP), can be interpreted
as the equilibrium state of an infinite-level fixed point iteration.
If we could construct a DL-based mapping whose fixed point corresponds
to the desired solution, as shown in the fourth column of Table
\ref{tab:Comparison-Existing-Proposed}, then the merits of classical
algorithms can be largely preserved. Iterative transceiver algorithms, e.g., AMP, contain a closed-form linear module and a non-linear module in each iteration. The latter is the performance bottleneck and can be replaced by NNs to enhance the performance. 

When designing such a mapping, it is imperative to consider two crucial factors. Firstly, the mapping should exhibit a rapid and stable convergence. Secondly, the converged fixed point should closely approximate the desired solution. The core idea of FPNs is to construct each iteration as a contraction mapping, i.e., with a Lipschitz constant smaller than one, to guarantee rapid convergence, and train the fixed point of this contraction mapping to closely match the desired solution. We will discuss how the proposed FPN framework addresses the challenges for
near-field XL-MIMO transceivers: 

\begin{itemize}
    \item \textbf{Theoretical Guarantee:} The Banach fixed point theorem suggests that applying a contraction mapping iteratively will monotonically converge to a unique fixed point in a linear rate \cite{2023Yu-JSTSP}. Since the Lipschitz constant of the linear module is readily available, one only needs to regularize the non-linear NN module to ensure that the concatenated mapping is contractive \cite{2023Yu-JSTSP}, which is easy to achieve and guarantees a fast convergence rate. 
    \item \textbf{Scalability and Complexity:} The training contains two steps, i.e., finding the fixed point iteratively and back-propagating through the fixed point. The implicit function theorem allows gradient calculation \textit{solely} with the fixed point, without storing any intermediate state \cite{2023Yu-JSTSP}. Hence, training incurs only a constant complexity, i.e., $O(1)$, in sharp contrast to DUNs, whose complexity increases with the number of layers $T$, i.e., $O(T)$. This is a big advantage as the huge system scale puts stringent requirements on complexity and scalability. The monotonic convergence also allows a flexible performance-complexity tradeoff. Since more iterations provably result in better performance, one can dynamically adjust the number of iterations according to the computational budget. 
    \item \textbf{Generalization:} Extensive empirical studies showed that FPNs have better generalization capability compared with DUNs under numerous distribution shifts \cite{2023Yu-JSTSP}. We explain this by FPN's insensitivity to initial values. Banach fixed point theorem indicates that the initial value of the fixed point iteration does not influence the unique fixed point of the contraction mapping. In addition, the small Lipschitz constant of contraction mapping can reduce the sensitivity to out-of-distribution perturbations. 
\end{itemize}
 
A case study on near-field XL-MIMO channel estimation is
provided in Section \ref{subsec:Channel-Estimation} to further demonstrate
the design guidelines of FPNs for practical problems. 

\section{Case Studies \label{sec:Case-Studies-for}}

\subsection{Beam Focusing \label{subsec:Beam-Focusing}}

Beam focusing is an important module for achieving high-rate multi-user
communications in near-field XL-MIMO \cite{2022Zhang}. The focused
beam is capable of providing high beamforming gains to combat the
severe propagation loss at mmWave/THz frequency bands and mitigate
the inter-user interference. 

\begin{figure}[t]
\centering{}\includegraphics[width=8.5cm]{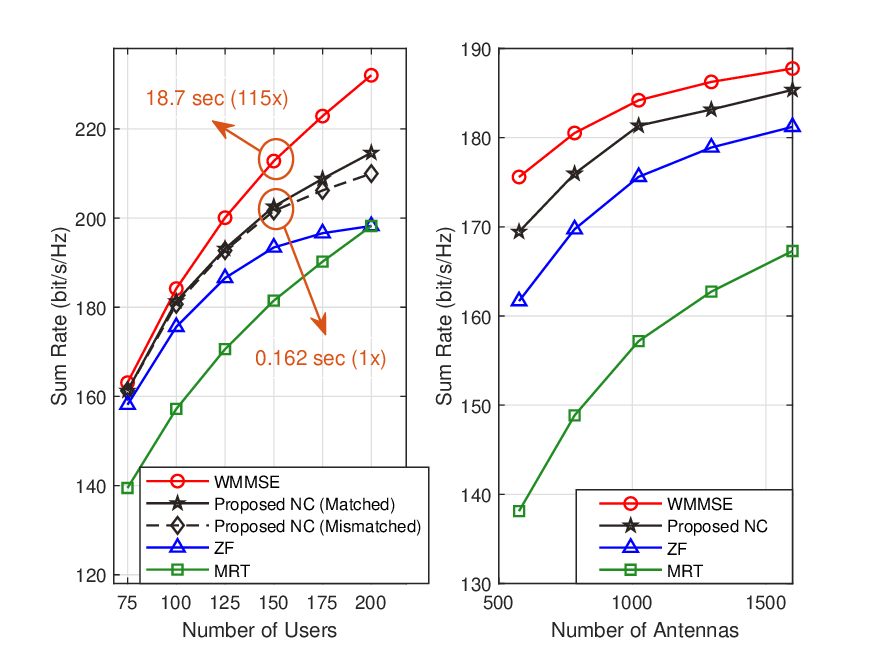}
\caption{(a) Sum rate versus the number of near-field users when the number of antennas is 1024. (b) Sum rate versus the number of antennas when the number of served users is 100. Settings: 14 GHz carrier frequency, 1 GHz system bandwidth, a fully-digital uniform planar array, -60 dBm noise power, 10 dB transmit power. A standard near-field channel model is adopted \cite{2023Cui}.} \label{fig:Sum-rate-as}
\end{figure}

For the downlink beam focusing problem, we can develop a scalable
and generalizable NC architecture by identifying the user-wise PE
property. The wireless network can be modeled as a directed
graph with node and edge features. In particular, the base station and each user are modeled as one node in the graph, and an
edge is drawn between the nodes if a transmission link exists. Then,
the beam focusing task can be formulated as an optimization problem
over the wireless channel graph and the user-wise PE property holds.
It indicates that the ordering of the users presented to the solver
should have no impact on system performance. This irrelevance of the
user ordering inspires us to construct $K$ duplicate CNNs that share
the same trainable parameters for $K$ users to learn the calibration
scheme \cite{2022Ma}. The unique NN structure is then combined with
the low-complexity ZF beamformer as the basis algorithm to achieve
high spectral efficiency as well as good scalability and generalization
ability. In such a way, the number of trainable parameters and the
training overhead are greatly reduced, which is crucial in near-field XL-MIMO. The resultant NC-based solver can also directly generalize when the number of users and their field types vary dynamically. 

Fig. \ref{fig:Sum-rate-as} demonstrates the
advantages of the proposed NC-based solver, where the classical
weighted MMSE (WMMSE) method is a performance upper bound
with extremely high complexity. It is observed from Fig. \ref{fig:Sum-rate-as}(a) that
the proposed NC-based method achieves at least 94.2\% performance
of WMMSE when the number of users is from 75 to 200, which reveals the high scalability in terms of served users. It is also shown in Fig. \ref{fig:Sum-rate-as}(a) that the iterative WMMSE algorithm requires 115 times longer runtime than the NC method when there are 150 near-field users, demonstrating the low complexity of NC. In Fig. \ref{fig:Sum-rate-as}(a), `matched' refers to the same number of users during training and inference, while `mismatched' represents the case where the training stage contains 75 users, but more than 75 users appear during inference. In addition, the proposed algorithm can generalize with respect to the number of users with a negligible drop in performance. This partially tackles the uncertainties in the propagation environments. By contrast, the black-box NN is not plotted as it fails to work in such a large-scale near-field system. It is also shown in Fig. \ref{fig:Sum-rate-as}(b) that the proposed NC framework scales well for different numbers of antennas. When the number of antennas varies from 576 to 1600, the proposed NC always outperforms conventional ZF and MRT beamformers by a large margin.

\subsection{Channel Estimation \label{subsec:Channel-Estimation}}

Channels play a pivotal role in wireless communications
and are particularly important for near-field XL-MIMO systems. Besides the system scale, the main difficulty
of channel estimation lies in the mixture of far- and near-field
paths, which hampers classical sparsity-based methods and motivates the application of DL frameworks. Specifically, we present an FPN-based estimator in combination with the Orthogonal AMP (OAMP) algorithm, called FPN-OAMP \cite{2023Yu-JSTSP}. 
\begin{figure}[t]
\centering{}\includegraphics[width=8.5cm]{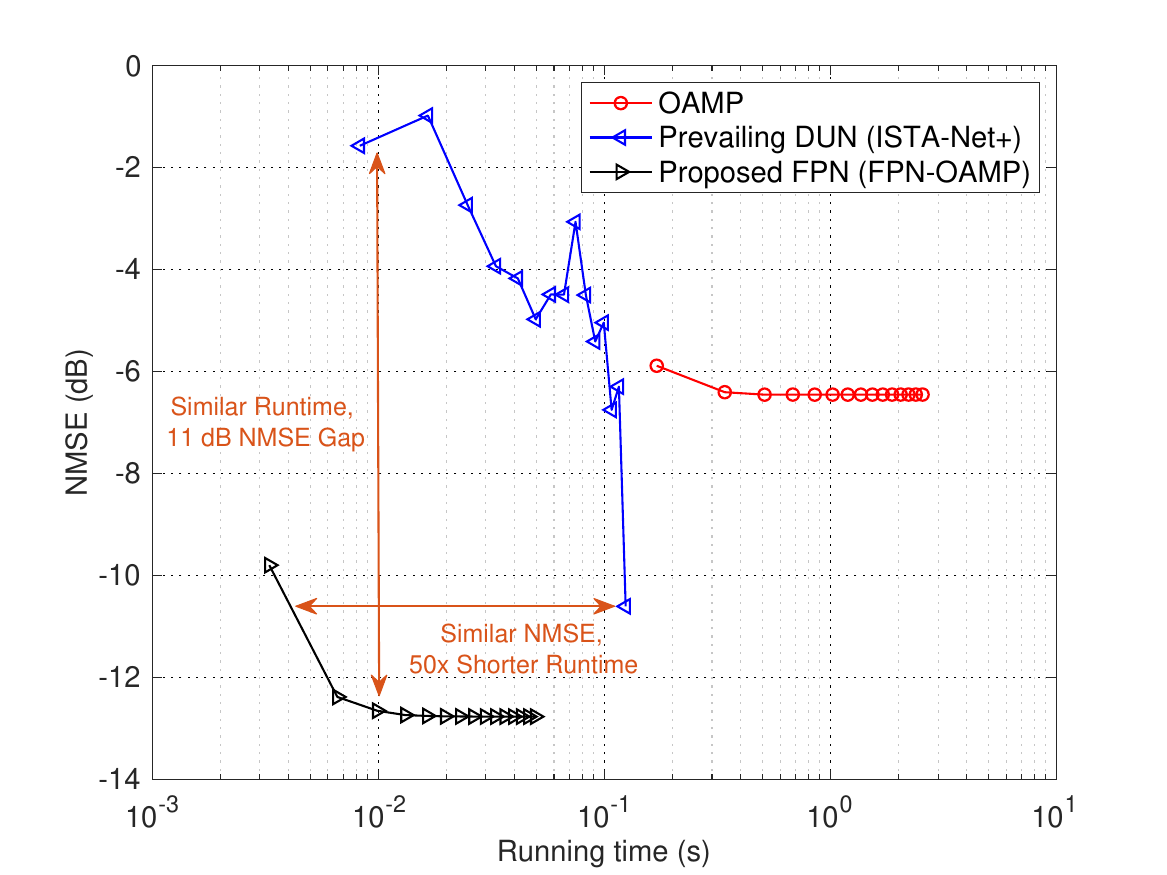}\caption{NMSE versus the CPU runtime. Settings: 300 GHz carrier frequency, a 1024-antenna non-uniform
array-of-subarray, 5 dB SNR, a one-bit hybrid analog-digital combiner with 50\% compression ratio. ISTA-Net+ is
the SoA DUN method for channel estimation. The hybrid-field
channel model in \cite{2023Yu-JSTSP} is adopted in the simulations.
\label{fig:NMSE-performance-as}}
\end{figure}

Each iteration of the OAMP algorithm consists of a linear estimator
(LE), which enforces consistency of the solution with the received
pilot signals, and a non-linear estimator (NLE), which encourages
the fidelity with the prior distribution of the XL-MIMO channel. The
main difficulties stem from the NLE, since the prior distribution
of XL-MIMO channel is hard to model, while the LE is purely dependent on the system model and the received pilots, which are readily available in
practice. Hence, we keep the LE intact and replace the NLE with a
specifically designed NN. The mapping in each iteration is a concatenation
of both. After analyzing the Lipschitz constant of the LE, we can
easily work out that of the NN in the NLE to ensure the contractiveness
of the overall mapping and achieve this during the training stage.
We adopt a residual CNN in the NLE thanks to its superb denoising
capability \cite{2023Yu-JSTSP}. It is worth noting that since the
training stage only requires storing the fixed point rather than every
intermediate solution, the space complexity is much smaller compared
to the DUN. Even low-end devices can complete the training
process for XL-MIMO systems, greatly reducing the deployment
cost. 
\begin{table}[t]
\caption{Out-of-distribution generalization performance of the proposed FPN-OAMP.
The setup is the same as Fig. \ref{fig:NMSE-performance-as}.
\label{tab:Out-of-distribution-generalizati}}

\centering{}\includegraphics[width=8.5cm]{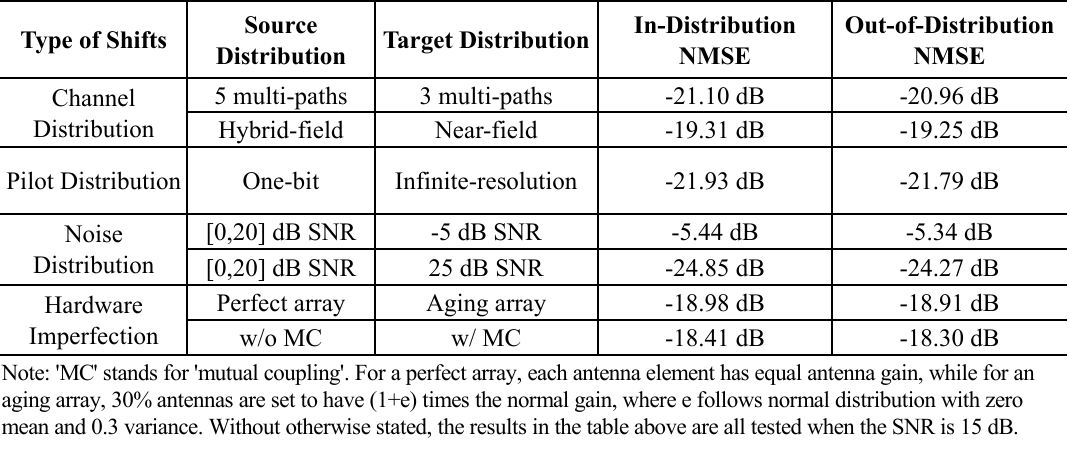}
\end{table}

Fig. \ref{fig:NMSE-performance-as} shows both
a significant speedup and a notable performance advantage of FPN-OAMP
compared with the SoA DUN method, i.e., ISTA-Net+. The proposed FPN achieves a similar NMSE with over 50 times shorter runtime compared to ISTA-Net+. While DUN fluctuates frequently, the proposed FPN exhibits a monotonic and stable convergence
towards the fixed point, allowing a flexible tradeoff between performance and complexity. In Table \ref{tab:Out-of-distribution-generalizati},
the out-of-distribution generalization capability of FPN-OAMP is
further demonstrated. Under practical distribution
shifts covering the channel, pilot pattern, noise, and hardware impairment, FPN-OAMP only exhibits a negligible performance decline, which verifies its robustness. 

\section{Future Directions and Open Issues}

While the proposed NC and FPN frameworks have demonstrated initial success in designing advanced near-field XL-MIMO transceivers, several open issues are worthy of further investigation in future research. 

\subsection{Unsupervised and Self-Supervised Learning}
Most existing works assume that the labels are perfectly known
and that the training dataset size is infinite (upon request) at the training stage. However, in extremely large-scale systems, these labels can
be hard, if not impossible, to obtain due to the prohibitive
computational cost (e.g., solving high-dimensional integer programs \cite{2022Ma}) and dynamic propagation environments
(e.g., shifts between far and near fields
\cite{2023Yu-JSTSP}). Studying unsupervised or self-supervised learning techniques that function without perfect labels is important, as they employ alternative loss functions to train the networks using noisy or corrupted raw signals and can alleviate the burden of label collection. 

\subsection{Co-Design with Advanced Neural Architectures}
We focused on presenting algorithm frameworks for designing DL-based transceivers, in which the learnable components are compatible with any neural network architecture. Nevertheless, the choice of the architecture can greatly affect the ultimate performance. For example, transformers can be more suitable to capture the energy leakage of near-field XL-MIMO channels than classical CNN architectures due to the effective attention mechanism \cite{2023Wang-transformer}. Recurrent architectures are more suitable than non-recurrent counterparts if the temporal correlation is apparent. In addition, graph neural networks can be combined to handle multi-user interference in XL-MIMO \cite{2024He,2022Kosasih}. Studying problem-specific choices of neural architectures is an interesting future direction.

\subsection{Enhanced Generalization Capability}
While the proposed NC and FPN frameworks have shown some ability to generalize across the different problem scales and data distributions, it is important to note that they are far from a universal solution. In future research, our proposal can be seamlessly integrated with numerous ongoing efforts aimed at augmenting the generalization capability beyond the inherent abilities of the algorithm frameworks, such as meta-learning, in-context learning, test-time training, etc. Combining the proposed frameworks with generative foundation models may also enable cross-domain generalization given effective prompts. 

\subsection{Problem-Specific Theoretical Guarantees}
While the proposed frameworks provide some theoretical supports, e.g., the convergence rates of FPNs and the PE property of NC, they are high-level properties from the perspective of algorithm frameworks. When it comes to specific problems and particular algorithms, stronger theoretical guarantees beyond these general properties can be derived. For example, it will be interesting to study the theoretical gap between the fixed point of FPNs and the optimal solutions of near-field channel estimation problems given some specific transceiver architectures. It is also worth studying the limit of the size generalization capability of NC for resource allocation problems in XL-MIMO systems under specific constraints.

\section{Conclusions}

In this article, we have analyzed the common challenges of near-field XL-MIMO transceiver design, and discussed how to utilize DL-based principles
to facilitate the transceiver design. We highlighted the challenges owing to system scale,
complex channels, and uncertainties in propagation environments. We proposed two DL frameworks for iterative and
non-iterative transceiver algorithms, respectively, that are
scalable, efficient, and robust. Two case studies for beam focusing
and channel estimation were analyzed to showcase the advantages of
these frameworks. Numerous future directions and open issues are also discussed to inspire further research.

\section*{Acknowledgement}
This work is supported by the Hong Kong Research Grants Council in part by the General Research Fund under Grant No. 16209023 and 16209622, and in part by the Areas of Excellence Scheme (Grant No. AoE/E-601/22-R). The work of Shenghui Song was supported by a grant from the NSFC/RGC Joint Research Scheme sponsored by the Research Grants Council of the Hong Kong Special Administrative Region, China and National Natural Science Foundation of China (Project No. N\_HKUST656/22). 

\bibliographystyle{IEEEtran}
\bibliography{references_WCM}

\noindent{\bf{Wentao Yu}} [S'21] (wyuaq@connect.ust.hk) is a Ph.D. candidate at HKUST, supervised by Prof. Khaled B. Letaief. \\
{\bf{Yifan Ma}} [S'19] (ymabj@connect.ust.hk) is a Ph.D. candidate at HKUST, supervised by Prof. Khaled B. Letaief. \\
{\bf{Hengtao He}} [M'21] (eehthe@ust.hk) received his Ph.D. degree from Southeast University. He is a Research Assistant Professor at HKUST. \\
{\bf{Shenghui Song}} [SM'21] (eeshsong@ust.hk) received his Ph.D. degree from the City University of Hong Kong (CityU). He is an Assistant Professor at HKUST. \\
{\bf{Jun Zhang}} [F'21] (eejzhang@ust.hk) received his Ph.D. degree from the University of Texas at Austin. He is an Associate Professor at HKUST. \\
{\bf{Khaled B. Letaief}} [F'03] (eekhaled@ust.hk) received his Ph.D. degree from Purdue University. He is the New Bright Professor of Engineering and Chair Professor at HKUST, and a Member of the US National Academy of Engineering. 

\end{document}